# Influence of extragalactic magnetic fields on extragalactic cascade gamma-ray emission

Anna Uryson

P.N. Lebedev Physical Institute of the Russian Academy of Sciences, Moscow, Russia

**Abstract**. We discuss the influence of extragalactic magnetic fields on the intensity of gamma-ray emission produced in electromagnetic cascades from ultra-high energy cosmic rays propagating in extragalactic space. Both cosmic rays and cascade particles propagate mostly out of galaxies, galactic clusters, and large scale structures as their relative volume is small. Therefore their magnetic fields weakly affect emission produced in cascades. Yet estimates of this influence can be useful searching for dark matter particles, when components of extragalactic gamma-ray background should be known, including cascade gamma-ray emission. To study magnetic field influence on cascade emission we analyze cosmic particle propagation in fields of $\sim 10^{-6}$ and $10^{-12}$ G (the former is typical inside galaxies and clusters and the latter is common in voids and outside galaxies and clusters). The calculated spectra of cascade gamma-ray emission are similar in the range of $\sim 10^7$-$10^9$ eV. So analyzing cascade emission in this range it is not necessary to specify models of extragalactic magnetic field.

e-mail: uryson@sci.lebedev.ru

## 1. Introduction.

Cosmic rays (CR) at ultrahigh energies (UHE) $E>4\times10^{19}$ eV are evidently of extragalactic origin. Propagating in space they interact with background emissions - cosmic microwave background (CMB) and radio emission via reactions: $p+\gamma_r \rightarrow p+\pi^0$, $p+\gamma_r \rightarrow n+\pi^+$. Pions and muons decay via $\pi^0 \rightarrow \gamma+\gamma$, $\pi^+ \rightarrow \mu^+ + \nu_\mu$, $\mu^+ \rightarrow e^+ + \nu_e + \bar{\nu}_\mu$ and neutrons decay producing p, $e^-$ and $\bar{\nu}_e$. Gamma-quanta, electrons and positrons are the particles which generate electromagnetic cascades in the interaction with CMB and radio background emission along with extragalactic background light (EBL): $e+\gamma_b \rightarrow e' + \gamma'$ (IC scattering) and $\gamma+\gamma_b \rightarrow e^+ + e^-$ (pair production). Other reactions give a minor contribution to electromagnetic cascades. The development of the electromagnetic cascades in the universe is described in [1-6].

In addition to processes above cascade electrons generate synchrotron radiation in the extragalactic magnetic field (EGMF), which quanta take part in IC scattering. This process affects the cascade process. The value of EGMF that weakly violate cascade development is estimated in [7]: $B<10^{-9}$ G.

EGMF is inhomogeneous, varying considerably in field strength: magnetic fields are $B\sim 10^{-6}$ G inside and near galaxies, in filaments and sheets fields are $B\sim 10^{-9}$-$10^{-7}$ G, and fields in voids are $B<10^{-11}$ G [8]. An experimental indication of EGMF is analyzed in, e.g., [9–12].

In this paper the EGMF effect on the intensity of diffuse cascade gamma quanta is discussed. It could be important searching dark matter (DM) particles, when any excess of gamma-ray background intensity is analyzed. The reason is that gamma quanta arise in decays of products of DM particle annihilation (see e.g. [13] and ref. therein). Thus the contribution of various components to the galactic background emission should be known, including cascade gamma-rays.

Regarding of the problem of components of extragalactic background emission, other models of gamma-ray production by UHECRs were also proposed [14-17]. In the model [14-16], it is suggested that UHE neutrinos born in UHECR interactions in extragalactic space interact with neutrinos in DM galactic halos and produce high-energy gamma-rays via the chain of reactions. Another possibility is that UHECRs are radioactive nuclei, the decay of which gives rise to electrons and consequent synchrotron gamma-rays [17]. However analyzing the contribution of these processes is beyond the scope of this paper.

At present, the structure and value of the EGMF is unclear. We suppose the EGMF to be uniform, its strength being a model parameter, and discuss cases $B\sim 10^{-6}$ and $B\leq 10^{-12}$ G.

The intensity of cascade gamma-rays was computed with the publically available code TransportCR [18] that simulates the propagation of UHE CRs and cascade particles.

We obtain that in the energy range of $\sim 10^7$-$10^9$ eV diffuse cascade emission is similar in the EGMF with $B=10^{-6}$ G and $B\leq 10^{-12}$ G. Thus analyzing diffuse cascade gamma-ray intensity at these energies specific to EGMF models is not necessary.

In the paper we use "photons" for particles of background emission, "quanta" for those produced in interaction with cosmic background and synchrotron process, and "electrons" for both electrons and positrons.

2. **Model.**

The model assumptions concern UHE CRs and their sources, extragalactic background emission, and EGMF structure and value.

We assume that CR sources are evidently of extragalactic origin, their sources being point-like objects, which are active galactic nuclei (AGN). Particle acceleration is connected with supermassive black holes in galactic centers, so AGNs possibly can be UHECR sources regardless of their type and distance [19, 20]. We suppose that CRs are accelerated on shock fronts in vicinity of supermassive black holes (e.g. in jets). This acceleration mechanism produces exponential injection spectra $\propto E^{-\alpha}$, with the spectral index $\alpha\approx 2.2$-$2.5$ [21-24]. Choosing the value of α we use results [24]. In this paper CR data including UHECR contribution to the galactic gamma-ray and neutrino backgrounds are described, varying values of injection spectral index α and exploring cosmological evolution of astrophysical objects where CRs possibly can be accelerated to UHE. Following [24] we adopt in the model that α=2.2.

The next question is supermassive black hole evolution. It is unclear, and in calculation we use the evolution of Blue Lacertae objects (BL Lac), which are one of the AGN types. The reason is that the bulk of CR data is described with it [24]. BL Lac's are located at distances corresponding to red shifts $z$≈0.0-5.

We assume that UHE CRs consist of protons.

Extragalactic background emissions are considered in the following way. The CMB has Planck energy distribution with the mean value $\varepsilon_r$=6.7×10$^{-4}$ eV. The mean photon density is $n_r$ =400 cm$^{-3}$. The background radio emission has parameters from the model of the luminosity evolution for radio galaxies [25]. The EBL parameters are taken from [26].

We suppose the EGMF to be uniform, its strength being a model parameter, and calculate cascade gamma-ray intensity with the code TransportCR [18] simulating particle propagation in extragalactic space.

### 3. Results and discussion.

Calculated gamma-ray spectra near the Earth in the fields of 10$^{-6}$, 10$^{-12}$ G are shown in Fig. 1. The spectra when $B$<10$^{-12}$ G coincide with that for $B$=10$^{-12}$ G, and are not shown in the figure. Varying parameters of EBL-model [26], the effect on the spectra is minor.

Now, we discuss reasons for the difference between the spectra. For a quick rough estimate we use the expression for the energy at which spectral distribution of the synchrotron emission of a single electron moving in the magnetic field has a maximum (see, e.g. [27]):

$$E_{s\,max} = 1.9 \cdot 10^{-20}\, H_{tr}\, (E_{eV})^2, \tag{1}$$

where $H_{tr}$ is the component of the field (in G) perpendicular to the electron velocity, $E_{eV}$ is the electron energy in eV. (We discuss here only the energy of maximum $E_{s\,max}$, as the flux of the synchrotron emission decreases with energy: it is several times lower at energies ≈10 times higher than $E_{s\,max}$). Synchrotron emission transfers electron energy in low-energy gamma quanta: using (1), in the EGMF of 10$^{-12}$, 10$^{-6}$ G an electron at the energy of 10$^{14}$ eV produces a bulk of quanta at $E_{s\,max}$=1.9·10$^{-4}$ and 1.9·10$^{2}$ eV respectively. The higher the field, the larger part of electron energy is outlaid on low-energy synchrotron quanta, thus the smaller part of energy is spent on cascade gamma-rays. As a result in the range of 10$^8$-10$^{14}$ eV, the gamma-ray intensity calculated with $B$=10$^{-6}$ G is lower than that with $B$=10$^{-12}$ G. At energies lower than 10$^8$ eV synchrotron quanta provide an excess of gamma-rays: the larger the field, the higher the excess. Thus, at energies $E<10^8$ eV, the curve with $B=10^{-6}$ G lies above the one with $B=10^{-12}$ G.

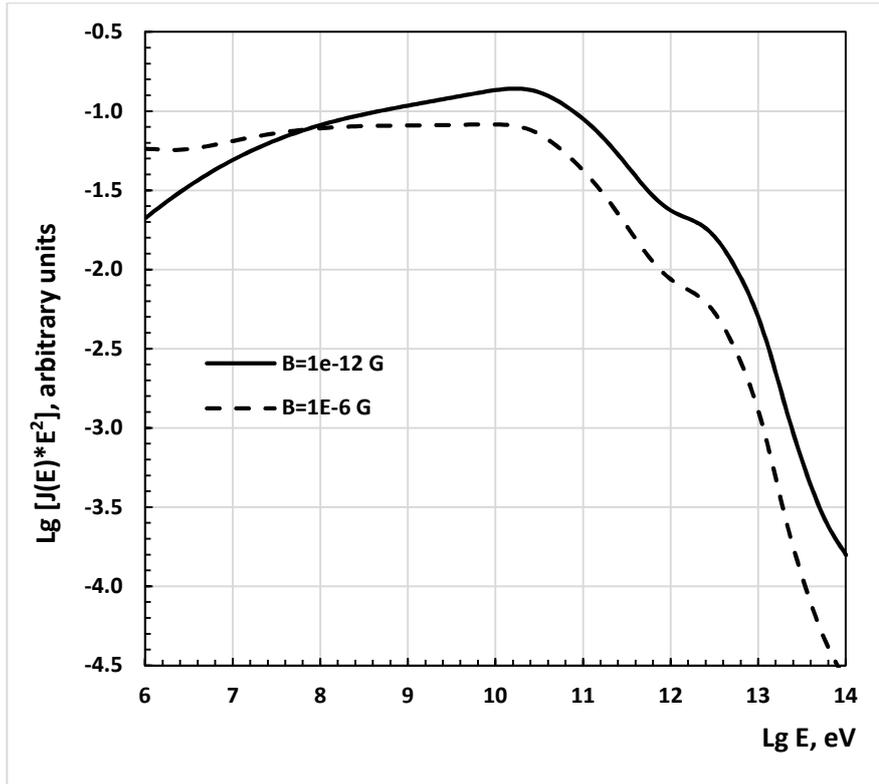

Fig.1. The spectra near the Earth of diffuse cascade gamma-rays produced by UHE protons in the extragalactic magnetic field $B$ of $10^{-6}$ and $10^{-12}$ G. The spectra when EGMF is $B<10^{-12}$ G and $B=10^{-12}$ G coincide and are not shown in the figure.

The energy value at which curves in Figure 1 intersect is governed by location of UHECR sources. This is illustrated in Figure 2, where the ratio $R$ of cascade gamma-ray intensity with $B=10^{-6}$ G to that with $B=10^{-12}$ G is shown. The log $R=0$ ($R=1$) corresponds to the intersection of curves in Figure 1.

As shown in Figure 2, the nearer the sources, the higher the energy at which log $R=0$. The biggest difference is between the curves in two extreme cases: for nearby sources with red shifts $z\approx0.001-0.003$ and remote sources with $z=4-5.5$.

The reason for the difference between the curves is the following. UHE CRs emitted from nearby sources give no rise to cascades, as free paths of electrons and quanta are longer than distances from the sources [28], so cascades do not develop. Electrons are born via $\mu^+$ and neutron decay, gamma-rays are produced only in $\pi^0$-decays and the synchrotron emission of electrons. $\pi^0$-decays produce few quanta [28], and the quanta main source is synchrotron emission. It depends on EGMF and as a result the intensity of gamma-ray emission (the most of which is synchrotron emission) is much higher in the EGMF of $10^{-6}$ G than that in the EGMF of $10^{-12}$ G, compared with the case of remote sources.

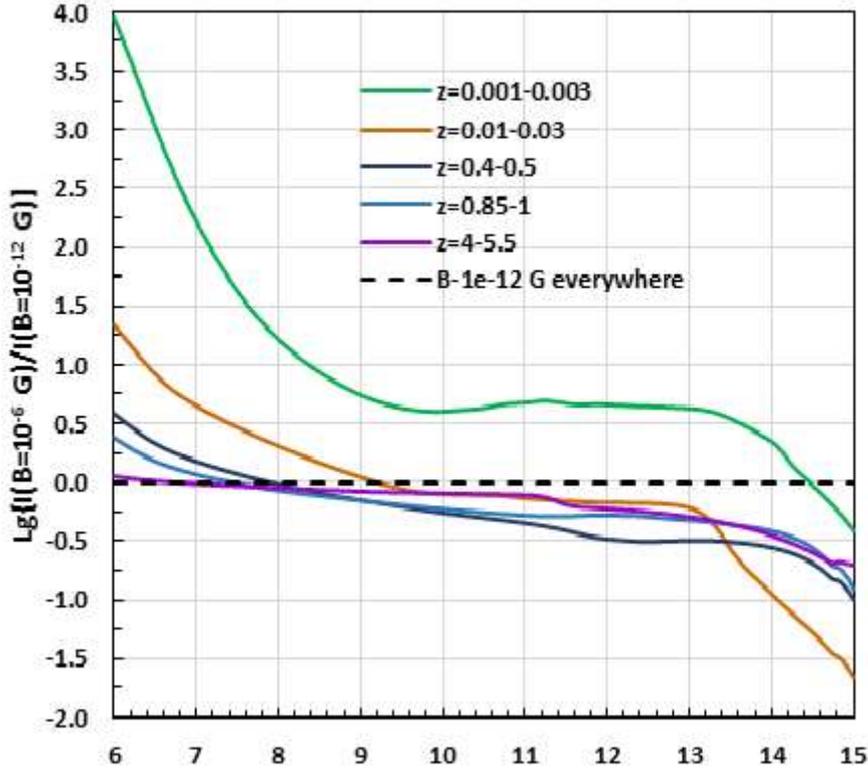

Figure 2. The ratio of cascade gamma-ray intensity with $B=10^{-6}$ G to that with $B=10^{-12}$ G for UHECR sources at various distances from Earth (see the legend).

In addition UHECRs from remote sources initiate cascades, in which electrons spend energy both in IC-scattering and synchrotron emission. Then a smaller part of electron energy is spent on synchrotron radiation compared to nearby sources, thus smaller part of cascade gamma-rays is of synchrotron origin. So, the larger the distances from sources, the weaker the EGMF influence. Additionally, energy dissipation in cascade leads to a loss of information about processes, in which cascade particles spend the energy [6], either in IC-scattering or synchrotron emission. Thus, the farther the sources, the weaker the difference between curves and the line corresponding to the case $B=10^{-12}$ G. The smallest difference is between this line and the curve for sources at $z=4$-$5.5$. In addition, in the case of remote sources, the synchrotron quanta energies being lower, the energy value at which $\log R=0$ shifts to smaller values. In the model the bulk of sources is located at $z \leq 2.5$ [24], setting curve crossing in Figure 1.

In the range of $\sim 10^7$-$10^9$ eV relative curve deviation from each other is minimal, being of 0.25-0.3. Closeness of curves is due to cascade 'universality' described in [6]. At higher energies (to $\sim 10^{19}$ eV) the deviation is of $\sim 10$.

In the range of $\sim 10^7$-$10^9$ eV the intensity of cascade gamma-ray emission is least dependent on EGMF. Thus at energies of $\sim 10^7$-$10^9$ eV it is possible to study contribution of cascade emission to the extragalactic gamma-ray background without specific models of extragalactic magnetic field.

### 4. Conclusion.

EGMF influences on diffuse cascade gamma-ray emission through synchrotron radiation of cascade electrons and IC scattering of synchrotron quanta. EGMF of $\sim 10^{-12}$ G and lower evidently has no influence on electromagnetic cascades from UHECRs. The higher magnetic field exists inside galaxies and galactic clusters, where it is of $\sim 10^{-6}$ G, and in sheets and filaments where it is of $\sim 10^{-7}$-$10^{-9}$ G. All these structures fill the insignificant part of the space and the EGMF effect on the cascade emission is minor. Yet the knowledge of this effect can be relevant to the search for dark matter particles, where the contribution of various components to the extragalactic background emission should be known. The reason is the following. Gamma-quanta arise in the decay of particles produced in DM particle annihilation. Thus any excess in gamma-ray background intensity is analyzed when searching DM particles. In view of this, components of the extragalactic background emission should be known, one of which is cascade gamma-rays.

For that reason we analyze the EGMF influence on the diffuse cascade gamma-ray spectra. We consider the model in which the extragalactic magnetic field is uniform, its strength being a model parameter, and compare cascade gamma-ray spectra in EGMF with $B = 10^{-6}$ and $B \leq 10^{-12}$ G. We obtain that the relative difference between the spectra is minimal, of 0.25-0.3, in the energy range of $\sim 10^7$-$10^9$ eV. The relative space volume occupied by field of $\sim 10^{-6}$ G being small, this difference is minor, thus in this energy range specific models of extragalactic magnetic field are not required to study contribution of diffuse cascade emission in the extragalactic gamma-ray background.

The result obtained is seemingly valid also for purely electromagnetic cascades, as was discussed in [29].

**Acknowledgements.**

The author thanks O. Kalashev for the discussion of the code TransportCR and extragalactic cascade features, T. Dzhatdoev for the discussion of extragalactic magnetic fields.